\documentclass[useAMS,usenatbib,aas_macros]{mn2e}

\usepackage{lscape}
\usepackage{graphicx}
\usepackage{colortbl}
\usepackage{amsmath}
\usepackage{natbib}
\usepackage{times}
\usepackage{aas_macros}

\newcommand{\msun}{~\mathrm{M}_{\odot}}
\newcommand{\msunyr}{~\mathrm{M}_{\odot}\,\mathrm{yr}^{-1}}
\newcommand{\Myr}{~\mathrm{Myr}}
\newcommand{\K}{~\mathrm{K}}
\newcommand{\kpc}{~\mathrm{kpc}}

\newcommand{\erg}{~\mathrm{erg}}
\newcommand{\cmmt}{~\mathrm{cm}^{-3}}
\newcommand{\zsun}{~\mathrm{Z}_{\odot}}
\newcommand{\Mbh}{~\mathrm{M}_{\rm BH}}
\usepackage{color}

\def\simpropto{\lower.2ex\hbox{$\; \buildrel \propto \over \sim \;$}}
\def\ltsim{\lower.5ex\hbox{$\; \buildrel < \over \sim \;$}}
\def\gtsim{\lower.5ex\hbox{$\; \buildrel > \over \sim \;$}}

\begin{document}

\title[FiBY: birthplaces of DCBHs]{The First Billion Years project: birthplaces of direct collapse black holes}

\author[B. Agarwal, et al.]{Bhaskar Agarwal,$^1$\thanks{E-mail:
agarwalb@mpe.mpg.de} Claudio Dalla Vecchia,$^{2,3,1}$ Jarrett L. Johnson,$^{4,1}$
\newauthor Sadegh Khochfar,$^{5,1}$, Jan-Pieter Paardekooper$^1$\\
\\
$^1$Max-Planck-Institut f{\"u}r extraterrestrische Physik,
Giessenbachstra\ss{}e, 85748 Garching, Germany\\
$^2$Instituto de Astrof\'isica de Canarias, C/ V\'ia L\'actea s/n, 38205 La Laguna, Tenerife, Spain\\
$^3$Departamento de Astrof\'isica, Universidad de La Laguna, Av. del Astrof\'isico Franciso S\'anchez s/n, 38205 La Laguna, Tenerife, Spain\\
$^4$X Theoretical Division, Los Alamos National Laboratory, Los Alamos, NM 87545, USA\\
$^5$Institute for Astronomy, University of Edinburgh, Royal Observatory, Edinburgh, EH9 3HJ}


\date{}
\pagerange{\pageref{firstpage}--\pageref{lastpage}} \pubyear{0000}
\maketitle

\label{firstpage}


\begin{abstract}
We investigate the environment in which direct-collapse black holes may form by analysing a cosmological, hydrodynamical simulation that is part of the First Billion Years project. This simulation includes the most relevant physical processes leading to direct collapse of haloes, most importantly, molecular hydrogen depletion by dissociation of H$_2$ and H$^-$ from the evolving Lyman-Werner radiation field. We selected a sample of pristine atomic cooling haloes that have never formed stars in their past, have not been polluted with heavy elements and are cooling predominantly \textit{via} atomic hydrogen lines. Amongst them we identified six haloes that could potentially harbour massive seed black holes formed via direct collapse (with masses in the range of $10^{4-6}\msun$). These potential hosts of direct-collapse black holes form as satellites are found within $15$ physical kpc of proto-galaxies, with stellar masses in the range $\approx 10^{5-7}\msun$ and maximal star formation rates of $\approx 0.1\msunyr$ over the past $5\Myr$, and are exposed to the highest flux of Lyman-Werner radiation emitted from the neighbouring galaxies. It is the proximity to these proto-galaxies that differentiates these haloes from rest of the sample.
\end{abstract}



\section{Introduction}

The observation of multiple high redshift quasars at $z>6$ \citep[e.g.][]{Mortlock:2011p447,Venemans:2013p3633} demands an explanation for the origin and growth of the (supermassive) black holes (BHs) fuelling quasar activity. The idea of the direct collapse of pristine gas in primordial haloes into BHs, i.e. direct-collapse black holes (DCBHs), is aimed at solving this problem by providing a physical mechanism to form seed BHs with masses $\Mbh \sim 10^{4-6}\msun$ \citep[e.g.][]{Eisenstein:1995p870}. The main advantage of the DCBH scenario is a much larger seed mass than what is expected from population III (hereafter, Pop~III) stellar remnants ($\sim 100\msun$), which makes it easier to grow to supermassive scales in a relatively short time \citep[e.g.][]{Volonteri:2008p1043}. During collapse the gas cloud must avoid fragmentation and lose its angular momentum in order to form a high density gas core. The core could ultimately results in a DCBH if it can accrete at a rate of $\sim 0.1$-$1\msunyr$ for $10^{5-6}\Myr$ \citep[e.g.][]{Latif:2013p2787}.

That said, a pristine, low spin, atomic cooling halo with a critically low H$_2$ fraction is the prerequisite for DCBH formation \citep{Bromm:2003p22}. Pristine gas is necessary as the injection of metals in the halo (e.g.~from neighbouring supernova driven winds) would lead to fragmentation (and eventually star formation) during the collapse process, as metals are effective coolants \citep[e.g.][]{Omukai:2008p113}. A low spin halo can allow for the efficient transport of angular momentum, thereby facilitating the accumulation of gas towards the central region \citep{Lodato:2006p375}. A low H$_2$ fraction ensures that the pristine gas cools mostly \textit{via} atomic hydrogen, which sets the minimum cooling temperature at $\sim 8000\K$, thereby raising the Jeans mass required for collapse to $10^5\msun$ at a density of $10^4\ \rm cm^{-3}$. The suppression of H$_2$ cooling can be attained through a high level of Lyman-Werner (LW) radiation as it can effectively dissociate H$_2$ into atomic hydrogen \citep{Shang:2010p33,WolcottGreen:2011p121}. Following the prerequisites outlined above, the DCBH can form through various channels: the supermassive star stage \citep{Begelman:1978p792,Begelman:2010p872,Hosokawa:2012p3508,Hosokawa:2013p3513}, the quasi-star stage \citep[e.g.][]{Begelman:2008p672}, or \textit{via} runaway collapse from a gas disc \citep{Koushiappas:2004p871,Lodato:2006p375}. 

To understand DCBH formation, one must first probe its plausibility, i.e.~the conditions required for a halo to qualify as a direct-collapse (DC) candidate.
\cite{Dijkstra:2008p45} (D08 hereafter) used Monte Carlo merger trees to predict the existence of such sites in the high-redshift Universe. They employed two-point correlation functions and halo mass functions and predicted a few DC sites per co-moving Gpc$^3$ volume. A recent study by \cite{Agarwal:2012p2110} (A12 hereafter) used a suite of semi-analytical models, on top of a cosmological N-body simulation, to predict the abundance of DC sites at $z\sim6$. Their model included tracking halo histories using merger trees and the spatial variation of LW radiation from both Pop~III and Pop~II stars. They predicted as many as few DC sites per co-moving Mpc$^3$, which is significantly higher than the earlier estimate (D08) and mainly due to taking into account the local variation in LW flux due to clustered star formation and the revision in the value of the critical level of LW radiation required to cause DC \citep{Shang:2010p33,WolcottGreen:2011p121}. Although A12 was an improvement over the earlier estimates of abundances of DC sites, the model was missing metal dispersion in the inter-galactic medium due to supernova driven winds, the self consistent treatment of gas physics (e.g. cooling, dissociation, photoionisation) and thereby could have been an overestimate.

Several hydrodynamical simulations have been employed to study the processes by which gas lose angular momentum and lead to the formation of a dense cloud that can undergo runaway gravitational collapse \citep[e.g.][]{Oh:2002p836,Bromm:2003p22,Begelman:2006p75}. Turbulence has been found to be one of the main agents via which gas can accumulate at the centre of metal-free atomic cooling haloes \citep{Wise:2008p2441,Latif:2013p2783}. However, the formation of a galactic-type disc has also been reported \citep{Regan:2009p776}. Note that, in all these studies, single, isolated haloes, extracted from cosmological simulations in some cases, were assumed \textit{a priori} to fullfill the criteria for DC.

The strength of the current study is the use of a hydrodynamical, cosmological simulation that self-consistently accounts for pair-instability (PISN) and core-collapse SN feedback from Pop~III and Pop~II stars (in the form of enrichment and energetic feedback), the self-consistent evolution of the global and local photo-dissociating LW radiation from both stellar populations, and the photo-ionisation of atomic ($\mathrm{H}^-+\gamma\rightarrow \mathrm{H} + \mathrm{e}$) and photo-dissociation of molecular ($\mathrm{H}_2+\gamma\rightarrow \mathrm{H} + \mathrm{H}$) hydrogen species. The advantage of such approach is that, for the selected candidate haloes, we know their formation history and the environment they live in.

The paper is organised as follows. We briefly describe the FiBY simulation used in this work and the modelling of LW radiation and self-shielding in section~\ref{sec.method}. The results of our study are presented in section~\ref{sec.results} where we discuss the nature of the DC sites, their merger histories and the nature of the galaxies in their local neighbourhood. The summary of the work and the discussion of the results are presented in section~\ref{sec.summary}.

\section{Methodology}
\label{sec.method}

\label{FiBY Simulation}

\subsection{FiBY Simulation}
\label{fiby sim}
The simulation used for this work is one out of the suite of the First Billion Years (FiBY) project (Khochfar et al. in prep., Dalla Vecchia et al. in prep.). Here, we briefly describe the code employed in the project and but we highlight the key features of the simulation we used \citep[see also][J13 hereafter]{Johnson:2013p2049}. 

A modified version of the smoothed-particle hydrodynamics (SPH) code \textsc{gadget} \citep{Springel:2001p10,Springel:2005p667} based on the version developed for the Overwhelmingly Large Simulations (OWLS) project \citep{Schaye:2010p2481}, was used in FiBY. The simulation was run with an equal number of gas and dark matter (DM) particles, $684^3$ each, in a box with side length of 4~cMpc. The mass of each DM (gas) particle is $m_{\rm DM}=6161\msun$ (initially, $m_{\rm gas}=1253\msun$), which allows us to resolve a minimum Jeans mass of the order of $10^5\msun$ with 100 gas particles \citep{Bate:1997p2447}. We used a \textsc{fof} halo finder \citep{Davis:1985p3262} together with the \textsc{subfind} algorithm \citep{Springel:2001p10,Dolag:2009p3306} to identify over-dense, self-bound haloes and sub-haloes. Merger trees were constructed on the \textsc{subfind} outputs using the prescription of \cite{Neistein:2012p3258}.

\subsubsection{Star formation and SN feedback}
\label{star formation and SNe}

Star formation is based on a pressure law designed to match the Kennicut-Schmidt relation \citep{Kennicutt:1998p143}, as discussed in \cite{Schaye:2008p2511}. The threshold density for star formation is set to $n=10\cmmt$ which is sufficient to account for LW feedback in pristine haloes (J13). At this threshold density and for a temperature $T=1000\K$, the Jeans mass is resolved with several hundred particles. Pop~III stars follow a Salpeter initial mass function (IMF) \citep{Salpeter:1955p861} with upper and lower mass limits at $21\msun$ and $500\msun$, and form in regions with metallicity $Z<10^{-4}\zsun$, with $\zsun=0.02$. Pop~II stars follow a Chabrier IMF \citep{Chabrier:2003p2503}, and form in regions with metallicity $Z\geq 10^{-4}\zsun$. For a discussion of the choices of IMF and critical metallicity for star formation, we refer the reader to J13 and \cite{Maio:2011p104}.

We model the feedback from both Pop~III and Pop~II (PI)SNe by injecting thermal energy into the inter-stellar medium (ISM) that surrounds the star particle \citep{DallaVecchia:2012p2528}. SN feedback is implemented stochastically by heating few gas particles (a mass comparable to that of the stellar population releasing the energy) to a temperature of $\sim 10^{7.5}\K$ in order to avoid over-cooling.
Core-collapse SNe release an energy of $10^{51}\erg$ per SN for Pop~II (Pop~III) stellar masses in the range $[8,100]\msun$ ($[21,100]\msun$). PISNe release an average energy of $3\times10^{52}\erg$ per SN for Pop~III stellar masses in the range $[140,260]\msun$. The first source of energetic feedback are Pop~III stars that end their lives as (PI)SNe, and their total energy is injected once per star particle when the age of the stellar population is the lifetime of a $140\msun$ star. PISN are also the first source of pollution of the IGM.

The enrichment of the ISM is modelled by assuming that Pop~II and Pop~III star particles are continuously releasing hydrogen, helium, and heavier elements into the surrounding gas. The released elements follow abundances computed in accordance with tabulated yields for types Ia and II SNe, and for asymptotic giant branch (AGB) stars. The approach employed here is similar to that of \cite{Tornatore:2007p3006} and is described in \cite{Wiersma:2009p2530}. The same technique is used for the enrichment from Pop~III stars except that the tabulated yields were computed for metal-free stars \citep{Heger:2003p23,Heger:2010p96}.

\subsubsection{Modelling of LW radiation in FiBY}
\label{LW modelling}

We model the LW radiation specific intensity, $J_{\rm LW}$ (in units of $10^{-21}\erg\,\mathrm{s}^{-1}\,\mathrm{cm}^{-2}\,\mathrm{Hz}^{-1}\,\mathrm{sr}^{-1}$), in the form of a mean background as well as a spatially varying radiation intensity depending on the local distribution of stellar sources. The mean free path of LW photons in the early Universe can be up to 10-20 times the length of our simulated box \citep[e.g.][]{Haiman:1997p86}. Therefore, in order to compute the background, we use the approach of \cite{Greif:2006p99} who estimated a spatially uniform LW background as function of stellar mass density and redshift. We modify their approach to express the background as a function of the star formation rate density (per co-moving volume), $\dot\rho_*$, at any given redshift,
\begin{equation}
J_{\rm LW, III}^{\rm  bg} \simeq 1.5 \left(\frac{1+z}{16}\right)^{3} \left(\frac{\dot\rho_{\rm *,III}}{10^{-3} {\rm M_{\odot}\,yr^{-1}\,Mpc^{-3}}}\right)\,{\rm ,}
\end{equation}
\begin{equation}
J_{\rm LW, II}^{\rm bg} \simeq 0.3 \left(\frac{1+z}{16}\right)^{3} \left(\frac{\dot\rho_{\rm *,II}}{10^{-3} {\rm M_{\odot}\,yr^{-1}\,Mpc^{-3}}}\right)\,.
\end{equation}
The LW background intensity, at each simulation time step, is estimated for each stellar population individually (see J13). We assume here that the escape fraction of LW photons from their host haloes is equal to unity. However it is likely that this is an over-estimate, given that some fraction of LW photons are absorbed before escaping into the IGM \citep[e.g.][]{Kitayama:2004p669,Ricotti:2001p1122}. Although we assume that the ISM and IGM are optically thin to LW photons, we compute the self shielding and dissociating rates of H$_2$ and H$^-$ molecules depending upon the local gas density (see J13). We only use stellar populations with ages $< 5$~Myr for computing the background and the spatial variation, as the majority of LW photons are emitted within this time interval owing to either the lifetimes of the most massive stars or the spectral energy distribution of the sources \citep{Schaerer:2002p21,Leitherer:1999p112}.

As shown in A12, the local variation of LW radiation can be up to 4-5 orders of magnitude higher than the global mean \citep[see also ][D08]{Ahn:2009p77}. We account for this variation at any given spatial point in our box by summing up the contribution of the all local sources that are less than $5$~Myr old by using the following formulation
\begin{equation}
J_{\rm LW, III}^{\rm local} = \sum_{i=1}^{N_{\rm *, III}}  15 \left(\frac{r_{i}}{1 \, {\rm kpc}} \right)^{-2} \left(\frac{m_{*, i}}{10^3 \, {\rm M_{\odot}}} \right)\,,
\label{eq.local_III}
\end{equation} 
and
\begin{equation}
J_{\rm LW, II}^{\rm local} = \sum_{i=1}^{N_{\rm *, II}} 3 \left(\frac{r_{i}}{1 \, {\rm kpc}} \right)^{-2} \left(\frac{m_{*, i}}{10^3 \, {\rm M_{\odot}}} \right)\,,
\label{eq.local_II}
\end{equation}
where for every $i^{\rm th}$ individual star particle of mass $m_{\rm *, i}$, $r_{i}$ is its distance from the point in physical coordinates and $N_{\rm *, III}$ and $N_{\rm *, II}$ are the total number of Pop~III and Pop~II star particles respectively. We define the net LW specific intensity as
\begin{equation}
J_{\rm LW, II} = J_{\rm LW, II}^{\rm local} + J_{\rm LW, II}^{\rm bg}\,,
\end{equation}
\begin{equation}
J_{\rm LW, III} = J_{\rm LW, III}^{\rm local} + J_{\rm LW, III}^{\rm bg}\,.
\end{equation}

\subsubsection{Self shielding and dissociation rates of H$_2$ and H$^-$}
\label{H2 chem}
Self shielding can greatly impact the overall dissociation rates of molecular hydrogen \citep[e.g.][]{Draine:1996p2556,Glover:2001p2561}. We follow the approach of \cite{WolcottGreen:2011p121} (WG11 hereafter) and employ a technique that depends only on the locally stored quantities of each SPH particle. The self shielding of H$_2$ molecules is computed as function of the local column density, $N_{\rm H_2}$, defined by the local Jeans length (see J13) as
\begin{equation}
N_{\rm H_2} = 2 \times 10^{15} \, {\rm cm}^{-2}  \, \left( \frac{f_{\rm H2}}{10^{-6}} \right) \left(\frac{n_{\rm H}}{10 \, {\rm cm}^{-3}} \right)^{\frac{1}{2}}  \left(\frac{T}{10^3 \, {\rm K}} \right)^{\frac{1}{2}}\,,
\end{equation}
where $f_{\rm H2}$ is the H$_{\rm 2}$ fraction, $n_{\rm H}$ is the number density of hydrogen nuclei, and $T$ is the gas temperature.  

For a given gas particle with temperature $T$ and column density $N_{\rm H_2}$, the self shielding factor can be defined as (see J13)
\begin{eqnarray}
f_{\rm ss}(N_{\rm H_2}, T) & = & \frac{0.965}{(1+x/b_{\rm 5})^{1.1}} + \frac{0.035}{(1+x)^{0.5}}  \nonumber \\
& \times & {\rm exp}\left[-8.5 \times 10^{-4} (1+x)^{0.5} \right]\,,
\end{eqnarray}
where $x$ $\equiv$ $N_{\rm H_2}/5\times10^{14}$ cm$^{-2}$ and $b_{\rm 5}\equiv b/10^{5}$ cm s$^{-1}$.  Also, $b$ represents the Doppler broadening parameter, 
which in case of molecular hydrogen can be formulated as
\begin {equation}
b \equiv (k_{\rm B}T/m_{\rm H})^{\frac{1}{2}}\,,
\end{equation}
which leads to
\begin{equation}
b_{\rm 5} = 2.9 \left(\frac{T}{10^3 \, {\rm K}}\right)^{\frac{1}{2}}\,.
\end{equation}
The above formulation allows us to parameterise self shielding, i.e. the factor by which the level of LW radiation seen by a gas particle is locally attenuated.

The reaction rate coefficients for H$_2$ and H$^-$ are computed following \cite{Shang:2010p33} and by accounting for the contributions from the global and local LW radiation flux.

\begin{figure}
\centering
\includegraphics[width=\columnwidth,]{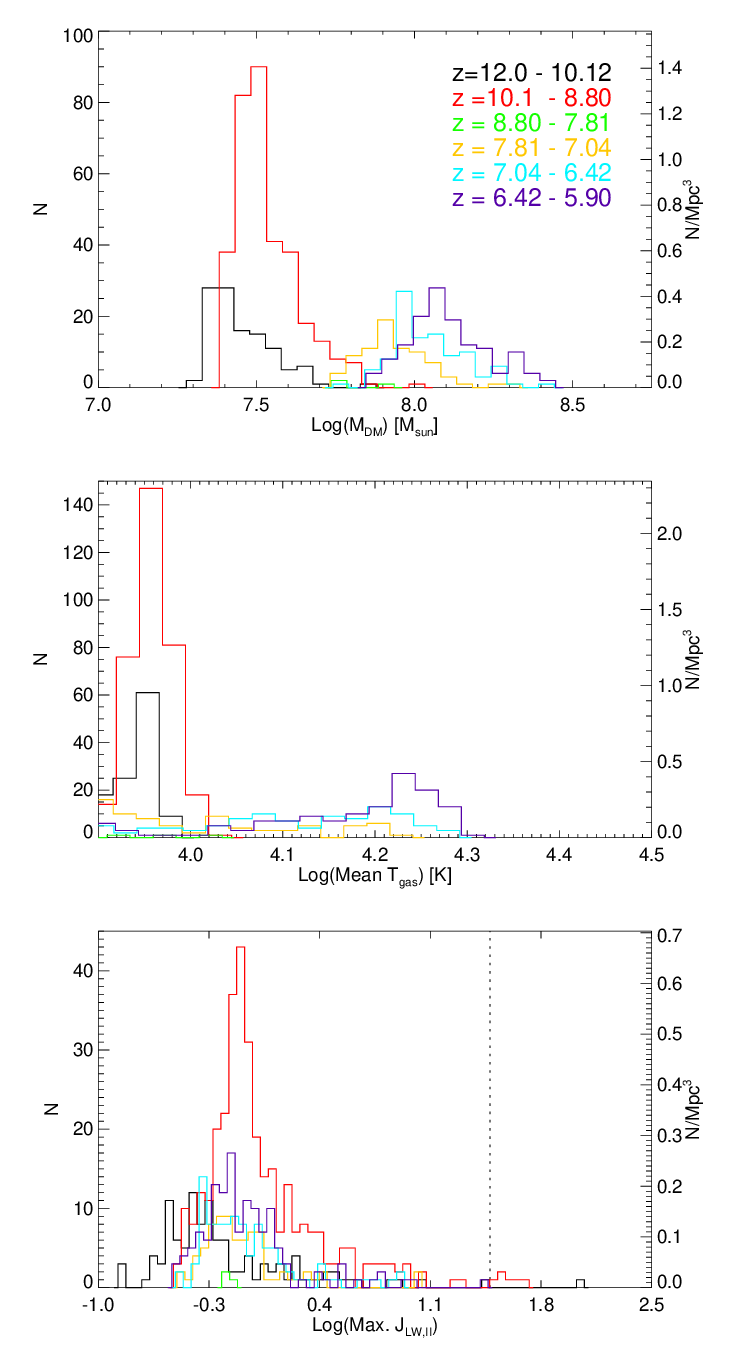}
\caption{Sample of pristine atomic cooling haloes with no star formation across all redshifts. \textit{Top, middle, bottom:} the haloes' DM mass, mass-weighted mean gas temperature and the maximum level of $J_{\rm LW, II}$ seen by a gas particle in a given halo. The dotted line denotes $J_{\rm LW, II}=30$.}
\label{fig.histograms}
\end{figure}

\section{Results}

We will now discuss the selection criteria used to identify pristine atomic-cooling haloes in the FiBY simulation analysed for this study. Here we mainly focus on the level of LW radiation seen by the candidate haloes, which is a crucial physical boundary condition for DC. Once the occurrence of the sites is explained, we will look into the evolutionary history of the haloes that are exposed to the highest levels of LW radiation. Finally, in order to understand their environment, we will discuss the nature of galaxies that produce the bulk of the LW radiation seen by these haloes.

\label{sec.results}
\subsection{Pristine, non star-forming, atomic cooling haloes}
\label{sec.dc selection}

We select pristine atomic-cooling haloes in which the gas satisfies the following conditions:
\begin{itemize}
\item virial temperature, $T_{\rm vir} \gtsim 10^{4}\K$. This condition ensures that only the haloes in which the gas predominantly cools by atomic H are selected, thereby satisfying one of the prerequisites for DCBH formation \citep{Oh:2002p836}. We use the relation from \cite{Barkana:2001p60} and compute the virial temperature of the halo based on its total mass, i.e. DM and gas, $M_{\rm halo} = M_{\rm DM} + M_{\rm gas}$. 
\item metallicity, $Z=0$, i.e. only the haloes that contain pristine (metal free) gas are chosen. This avoids fragmentation of the collapsing gas and eventual star formation. Note that this process may not be captured with the current simulation resolution.
\item stellar mass, $M_*=0$, throughout the history of the halo. This condition is imposed to make sure that the halo did not form any stars, nor any stars were accreted during eventual mergers.
\item star formation rate, $\dot\rho_*= 0$, throughout the history of the halo. Due to the stochasticity of star formation in the simulation, the criterion $M_*=0$ is not sufficient for selecting star-free haloes. Note that we can only check both conditions at simulation output times, where we also compute the haloes catalogues. However, the frequency of the outputs is large and the (average) time between simulation snapshots, is shorter than the dynamical time of the considered haloes.\footnote{The dynamical time of a halo defined as the ratio of the virial radius to the circular velocity, $t_{dyn} = R_{vir}/V_{cir}$, is 15 and 150 Myr at $z=30$  and $6$ respectively, whereas the simulation snapshots are timed at 6 and 53~Myr at the same epochs.}
\item {\bf halo contains at least 100 gas particles. This condition ensures that we exclude the numerically spurious haloes.}
\end{itemize}

\begin{table}
\caption[Properties of the DC candidate haloes]{\bf{Properties of the DC candidate haloes identified in the sample: haloes' DM mass, $M_{\rm DM}$, gas mass, $M_{\rm gas}$, redshift, $z$, the averaged LW specific intensity from Pop II stars, $J_{\rm LW,II}$, the averaged gas temperature and virial temperature. The averaged quantities are weighted by the mass of the gas particles in the halo.}}

\begin{center}
\begin{tabular*}{\columnwidth}{@{\extracolsep{\fill}}ccccccc}
 \hline \\ [-1.5ex]
 & $M_{\rm DM}$& $M_{\rm gas}$& $z$ & $J_{\rm LW,II}$ & T$_{\rm gas}$& T$_{\rm vir}$\\
 & $[10^7 \msun]$& $[10^6\msun]$& &[see text]&[K]&[K]\\
 \\ [-1.5ex] \hline \\ [-1.5ex]
DC0 & $2.37$ & $4.54$ & 10.49 & 145.90 & 8890 & 10181\\ 
DC1 & $2.65$& $3.93$ & 10.49 & 31.56 & 8592 & 10704\\
DC2 & $3.13$& $4.03$ &  9.65 & 33.81 & 9501 & 10963\\ 
DC3 & $3.25$& $5.42$ &  9.65 & 38.75 & 9877 & 11499\\ 
DC4 & $4.11$& $6.63$ &  9.25 & 69.97& 9234 & 12903\\ 
DC5 & $3.27$& $6.75$ &  8.86 & 47.09& 5658 & 10932\\
\hline
\end{tabular*}
\end{center}
\label{tab.1}
\end{table}

We plot the properties of the haloes that constitute our sample based on above criteria in Fig.~\ref{fig.histograms}, where the differently coloured histograms are spaced uniformly over time bins of 100 Myr. The lower mass cut off increases with decreasing redshift due to the fact that a fixed virial temperature cut-off of $10^4$~K implies a higher halo mass with decreasing redshift. The spread in the halo masses gets larger towards lower redshifts and the mean temperature of the gas in the haloes is highest in the lowest redshift bin, which is after the simulation volume has been re-ionised. 
{\bf The systematic increase in the mass (or virial temperature) of candidate haloes at lower redshift is due to the fact that some of the haloes that fulfil our selection criteria at high redshifts continue to grow in mass without forming stars and are as such included in our histograms at later times as well.
The rise in the gas temperature is not surprising, and follows the rise in the virial temperature of the haloes.}
The peaks of the histograms in the bottom panel do not necessarily reflect the background value of $J_{\rm LW, II}$ (see J13) in the simulation box in the corresponding redshift range. This is because the LW radiation level is expected to be higher than the global value, thereby providing a strong negative LW feedback and prohibiting Pop III star formation \citep{Machacek:2001p150,OShea:2008p41}.

\subsection{The DCBH candidate haloes} 

We define the LW specific intensity that a halo is exposed to as the average of the incident LW specific intensity over all particles within the virial radius (self-shielding is not included in this calculation). The critical value of LW radiation, $J_{\rm crit}$, refers to the specific intensity at which molecular hydrogen cooling is completely suppressed due to photodissociation of H$_2$ into H \citep{Omukai:2001p128}. This value depends on the underlying stellar population and is $J_{\rm crit,III}=1000$ from Pop~III stars \citep{WolcottGreen:2011p121} and $J_{\rm crit,II}=30$ from Pop~II stars \citep{Shang:2010p33}. In order to refine our criteria, we identify the haloes that are most likely to be DC candidate sites as the haloes exposed to $J\geq J_{\rm crit}$, and as soon as they cross the atomic cooling limit in their respective histories. {\bf{In this study, we choose $J_{\rm crit}=30$ which is the lower limit for the critical intensity. A recent study \citep{Latif:2014p3556} has pointed out the implications of choosing a slightly higher $J_{\rm crit}$ threshold which leads to a drop in the number density of such objects by almost 3 orders of magnitude (A12). Although the number of DC sites in our simulation volume is sensitive to this choice, we chose the lower limit to get a handle on the various environments that such sites might exist in.}} We thus identify 6 such potential DCBH hosts and refer to them as DC0-DC5. The haloes' DM and gas masses, the redshift at which they are first identified as DC candidates, and the LW specific intensity for the candidates is listed in Table~\ref{tab.1}. For the candidate sample, we find that $J_{\rm crit,II}$ is exceeded for all haloes, whereas $J_{\rm crit,III}$ is never reached.\footnote{This value for $J_{\rm crit,II}$ is likely a lower limit due to the authors' adoption of a stellar surface temperature of $10^4\rm \ K$ \citep{Shang:2010p33} , which is likely too low for Pop II stellar populations with ages less than 1~Gyr in the early universe. We note, however, that we have identified candidate haloes which are exposed to fluxes exceeding $J_{\rm LW, II}=100$, which is well above this minimum value.} We also find that the average H$_2$ fraction in these haloes, $f_{\rm H_2}< 10^{-7}$, well below the critical value at which molecular hydrogen cooling is completely suppressed \citep{Shang:2010p33}.

\begin{figure}
\centering
\includegraphics[width=\columnwidth,]{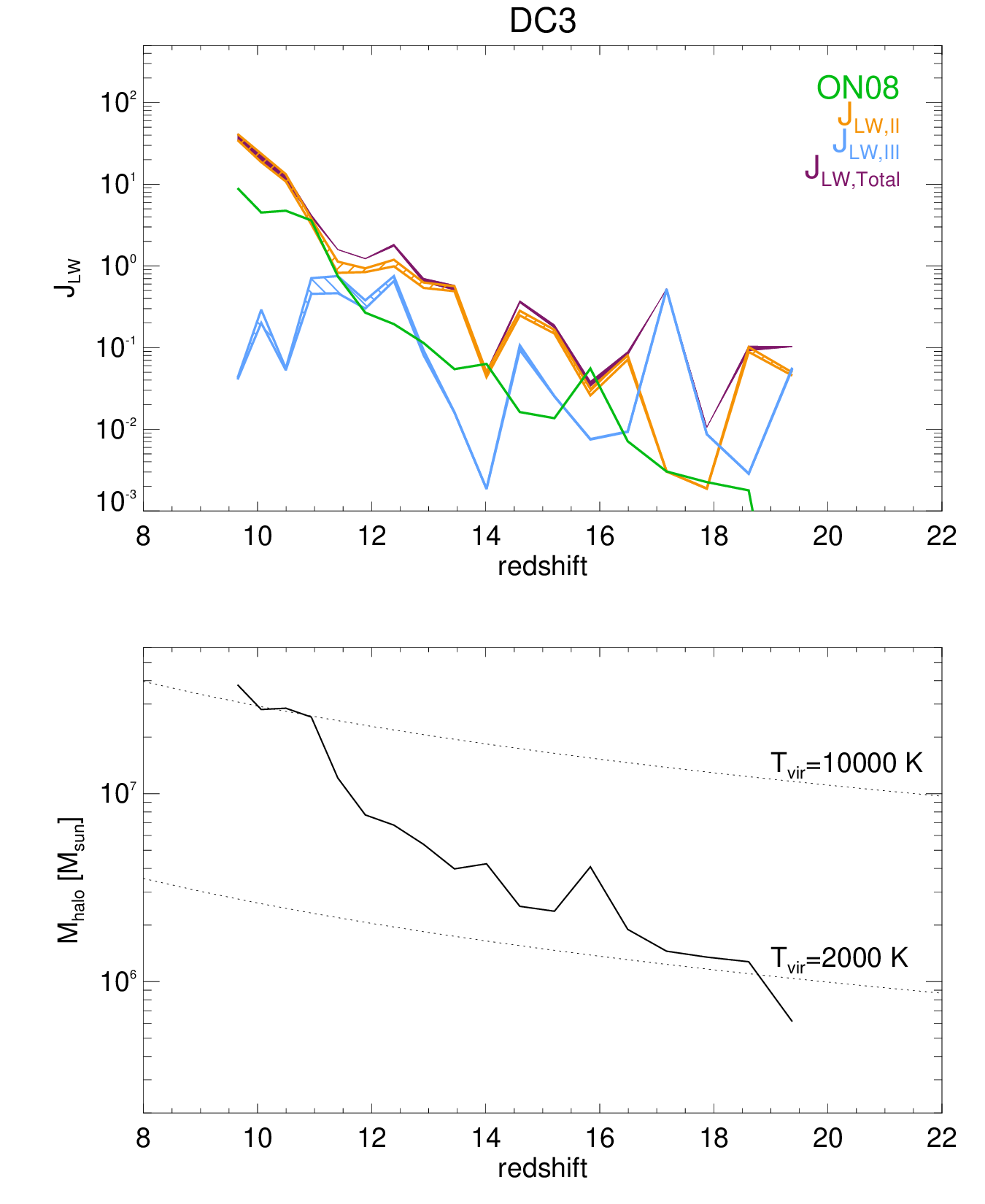}
\caption[Evolution of LW radiation in DC3]{{\textit {Top Panel}}: The $J_{\rm LW}$ radiation as seen by DC3 in its past. We track the DC halo all the way to its \textit{birth} (i.e. the first time it appears in the simulation volume) using merger trees and plot the maximum and minimum value (resulting in the spread and thus the shaded region) of the local LW radiation seen by the particles in the halo at each epoch. The $J_{\rm LW,III}$ is shown in blue, $J_{\rm LW,II}$ is shown in dark-yellow, the total is shown in dark-red and the green line denotes the level of LW specific intensity required by the halo at the given redshift to host a Pop III star (Eq. 2 in ON08). \textit{Bottom Panel:} The mass of the halo (DM + gas) plotted against redshift. The dotted lines correspond to the limits of $T_{\rm vir}=2000$ and $10^4 \ \rm K$ at each redshift.}
\label{fig.dc3_jlw}
\end{figure}

\begin{figure}
\centering
\includegraphics[width=\columnwidth,]{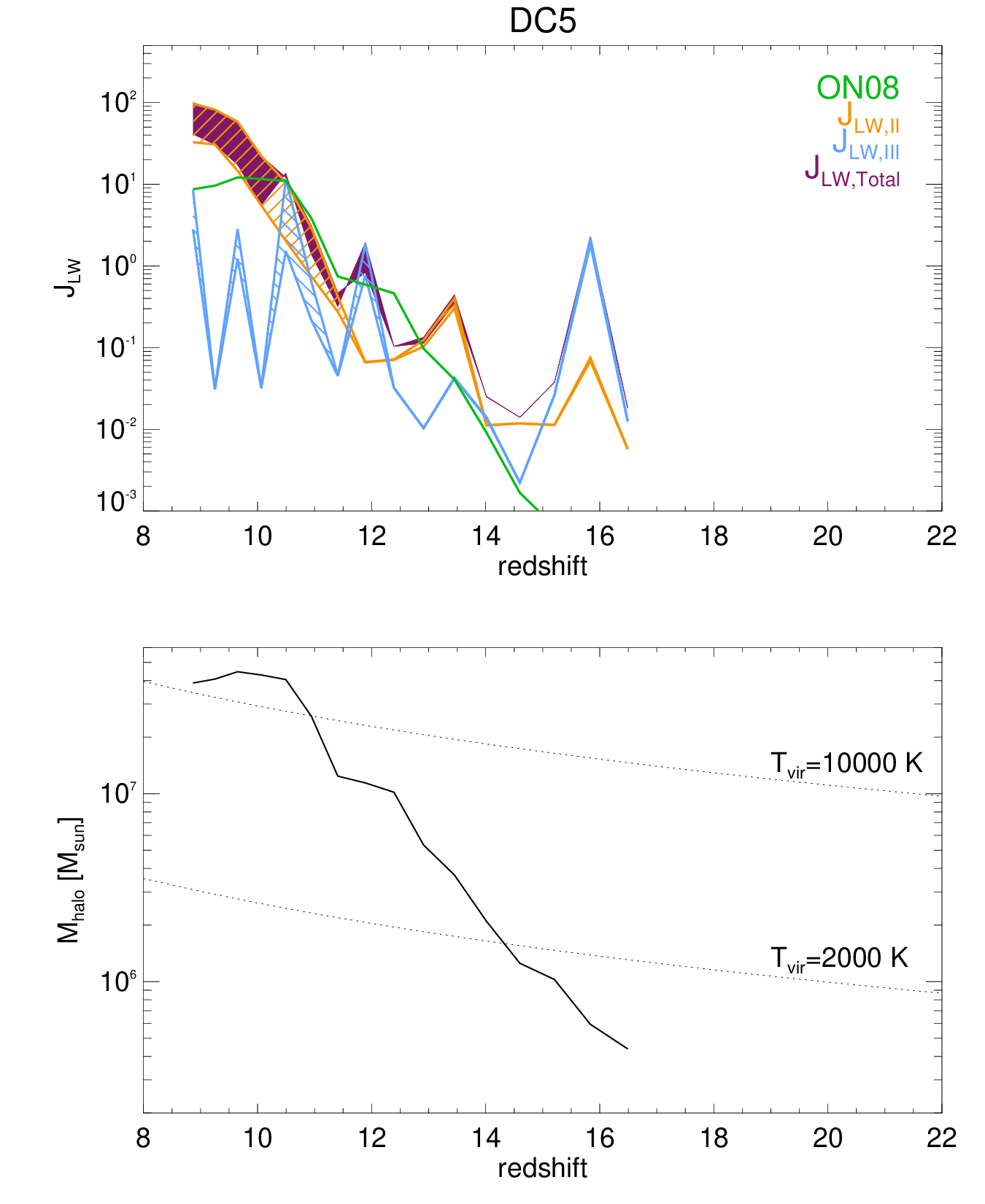}
\caption[Evolution of LW radiation in DC5]{Same as in Fig. \ref{fig.dc3_jlw} except for DC5.}
\label{fig.dc5_jlw}
\end{figure}

We compute the minimum and maximum value of the net LW flux that every gas particle in the halo receives at every snapshot before the halo reaches the virial cooling threshold. We do this by using merger trees and simulation snapshots: we follow the halo's history and identify it at each snapshot time; we then compute the LW flux for seen by all the gas particles in the halo at that time using the simulation output. This net LW specific intensity is a sum of the global and the local values and will determine whether or not the halo can host a Pop III star. This effect arises due to the dissociation of a fraction of H$_2$ molecules in the pristine haloes, which makes H$_2$ cooling less efficient and thus causes a delay in Pop~III star formation \citep[ON08 hereafter]{OShea:2008p41}. We plot the corresponding LW flux history of DC3 and DC5 in the top panels of Figs.~\ref{fig.dc3_jlw} and \ref{fig.dc5_jlw}. The evolution of the halo mass is plotted in the bottom panels. The net level of LW radiation (top panels, in maroon), is higher than the value that would allow for efficient H$_2$ cooling (in green, ON08). Note that, at early times, Pop~III stars are able to dominate the overall LW flux seen by the halos, but the trend quickly reverses as soon as Pop~II stars are able to form efficiently, thereby inhibiting Pop~III star formation in neighbouring mini-haloes. The exposure of the pristine haloes to LW radiation \textit{throughout their lifetimes} plays a critical role in allowing them to host DCBHs at later times, which is demonstrated in Fig.~\ref{fig.dc3_jlw} and \ref{fig.dc5_jlw}. The importance of both Pop~II stars (at later times) and Pop~III stars (at early times) is highlighted in the plot, where the combined specific intensity from both the stellar populations is imperative in suppressing Pop~III star formation in pristine haloes.

\subsection{The environment of the haloes hosting DCBHs}

\begin{figure*}
\centering
\includegraphics[trim = 15mm 15mm 2.5mm 0mm, clip,width=0.6\columnwidth,]{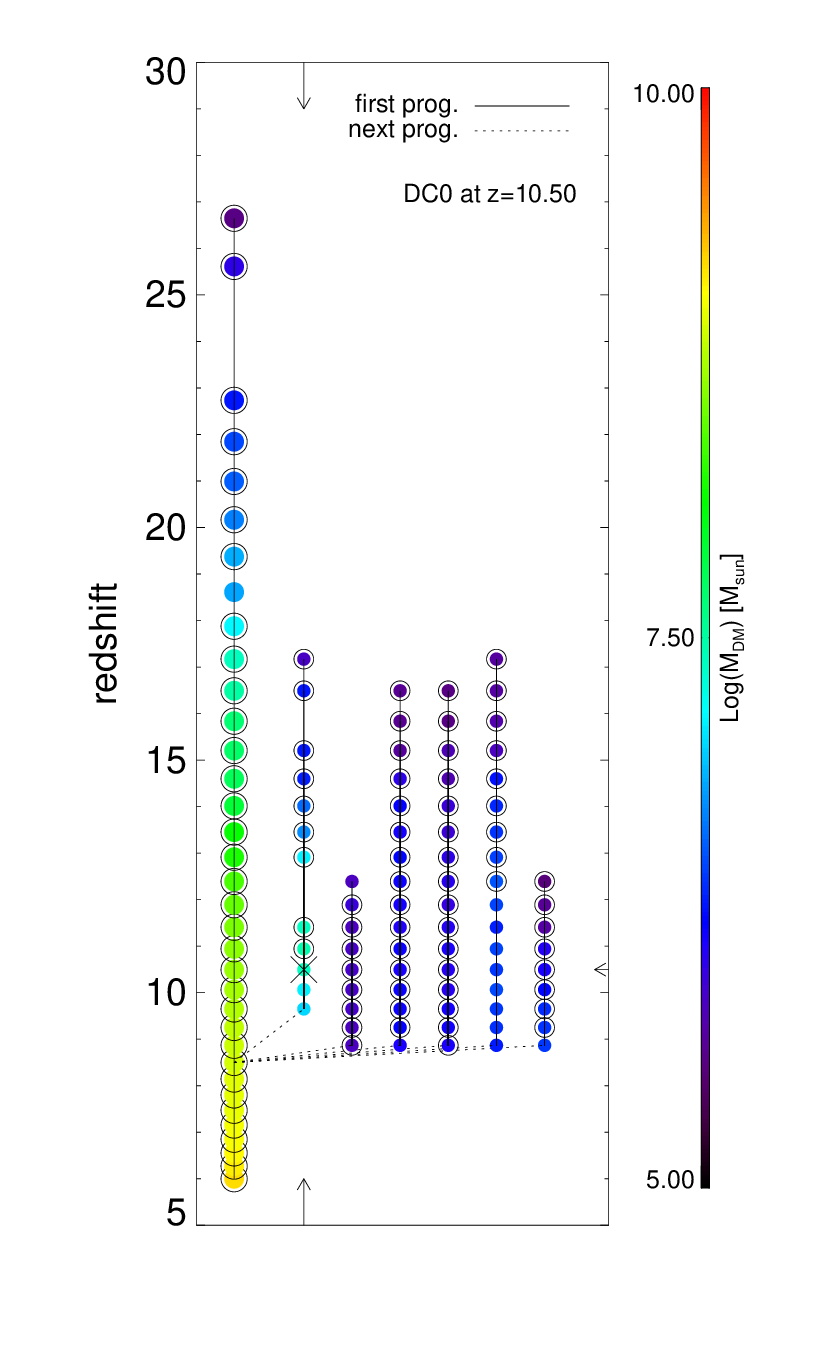}\includegraphics[trim = 15mm 15mm 2.5mm 0mm, clip,width=0.6\columnwidth,]{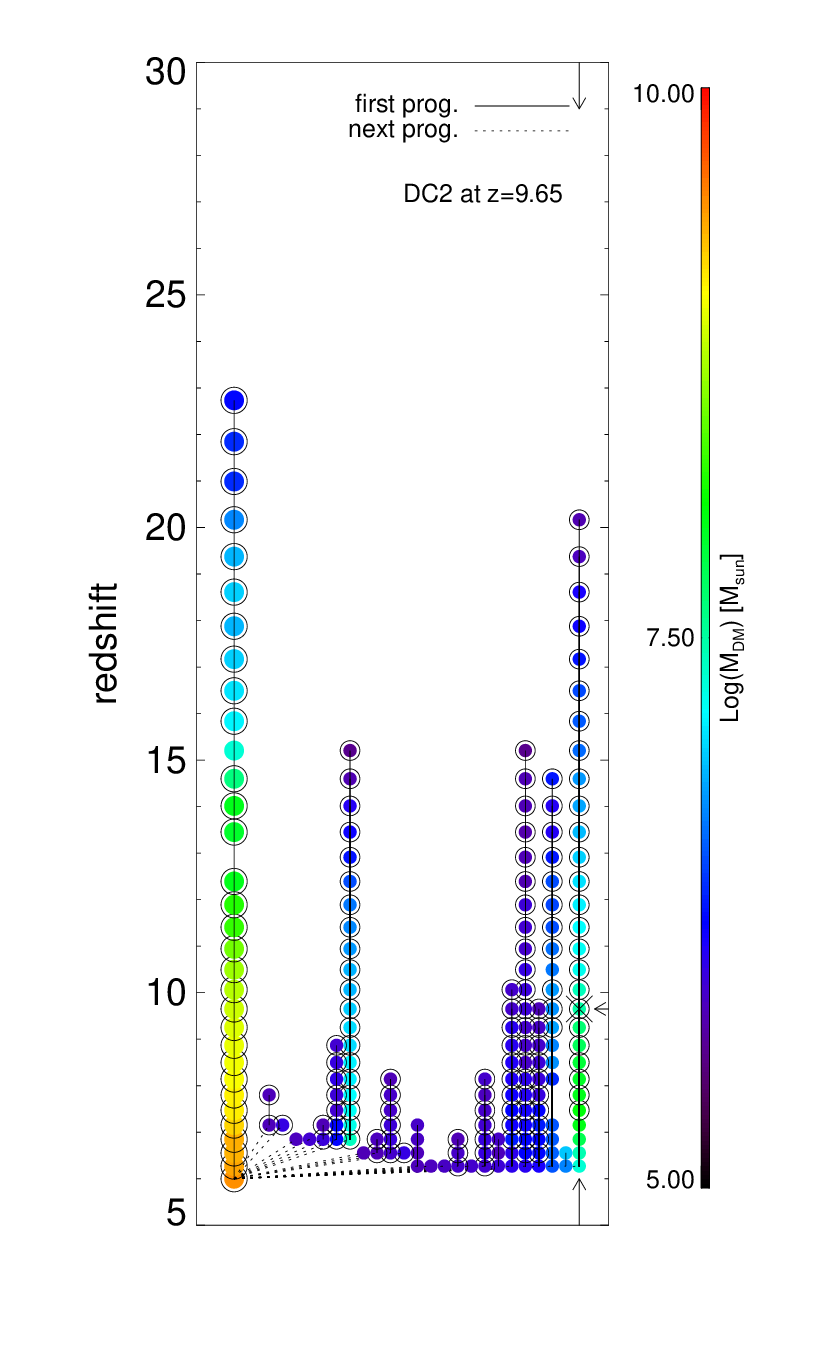}\includegraphics[trim = 15mm 15mm 2.5mm 0mm, clip,width=0.6\columnwidth,]{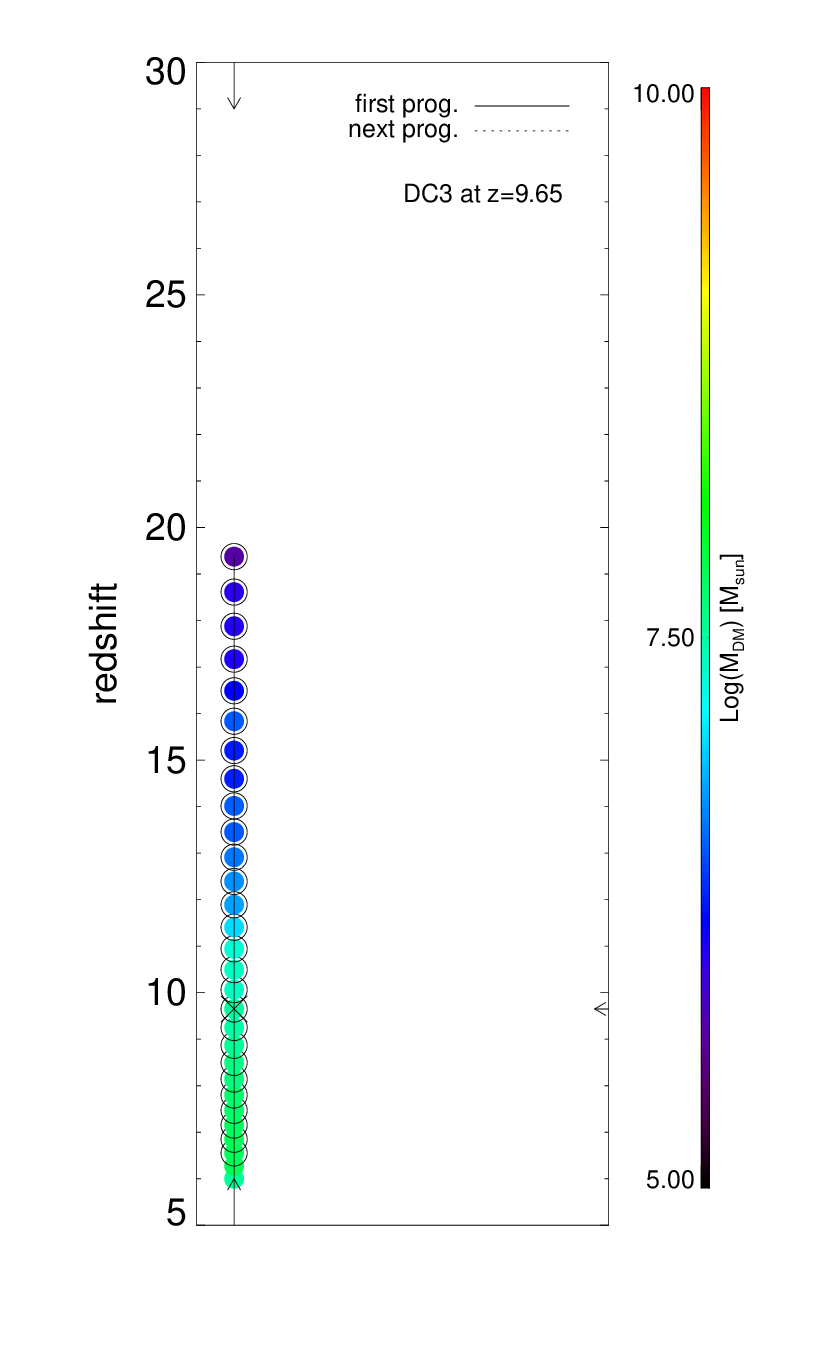}
\caption[Merger tree for DC candidate haloes.]{The merger tree for the DCBH candidate haloes, DC0, DC2 and DC3 marked by the cross (use arrows to guide the eye). The left-branch represents the main progenitor branch of the halo with which the DC candidate merges. On the right of this branch, we plot the merger history of the main progenitor halo with which the DC candidate merges. Enclosing circles imply that the halo is the most massive halo within its FoF group. The haloes are colour-coded by their DM mass.}
\label{fig.mergertrees}
\end{figure*}

\begin{figure*}
\centering
\includegraphics[trim = 0mm 0mm 0mm 15mm, clip,width=1.75\columnwidth,]{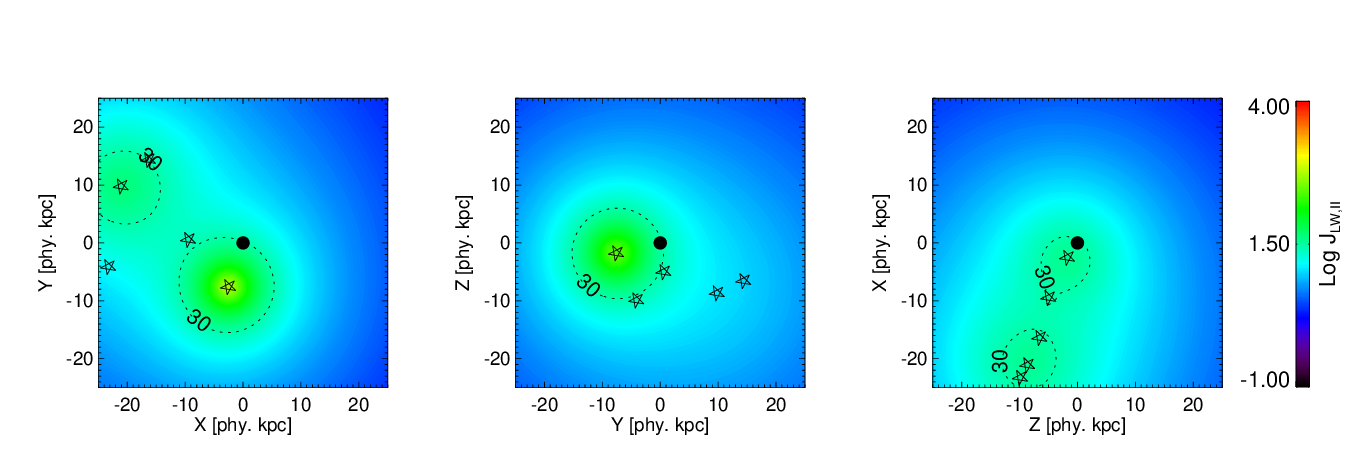}\\
\includegraphics[trim = 0mm 0mm 0mm 15mm, clip,width=1.75\columnwidth,]{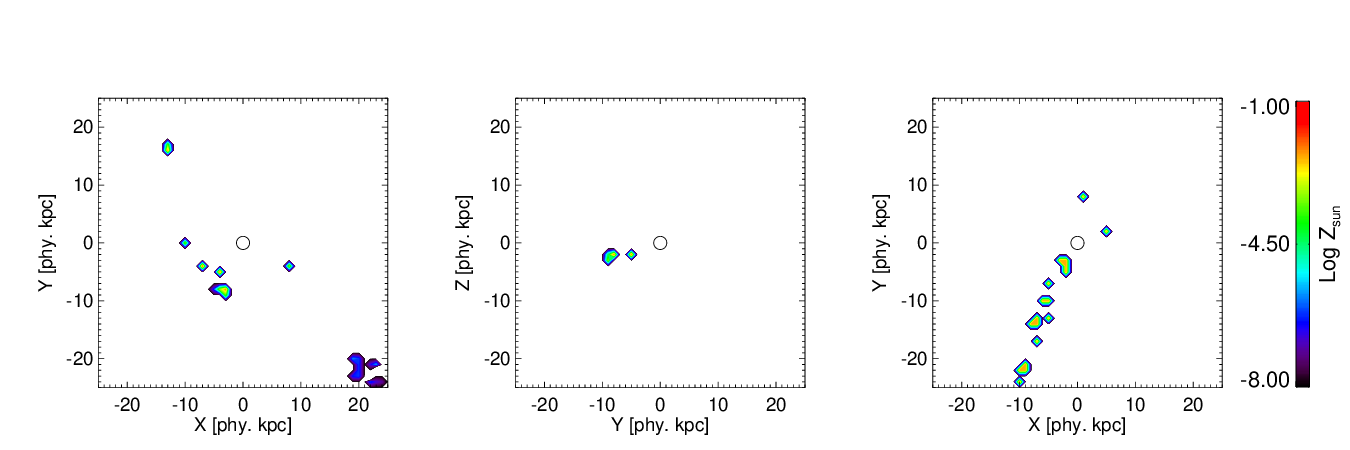}
\caption[LW radiation and metallicity for DC0's environment]{{\textit {Top}}: $J_{\rm LW, II}^{\rm local}$ slices along the XY, YZ and ZX plane centred around DC0 (black dot) spanning 50 physical kpc along each axis. The neighbouring galaxies producing the $J_{\rm LW, II}^{\rm local}$ are marked by a star symbol. The dotted contour line marks the region enclosing $J_{\rm crit,II}=30$. \textit{Bottom}: Metallicity slices along the XY, YZ and ZX plane centred around DC0 (empty circle) spanning 50 physical kpc along each axis and averaged over 10 physical kpc over the remaining axis. Interestingly despite the close proximity of galaxies, the DC candidate exists in a metal free region.}
\label{fig.DC0 slices}
\end{figure*}

\begin{figure*}
\centering
\includegraphics[trim = 0mm 0mm 0mm 15mm, clip,width=1.75\columnwidth,]{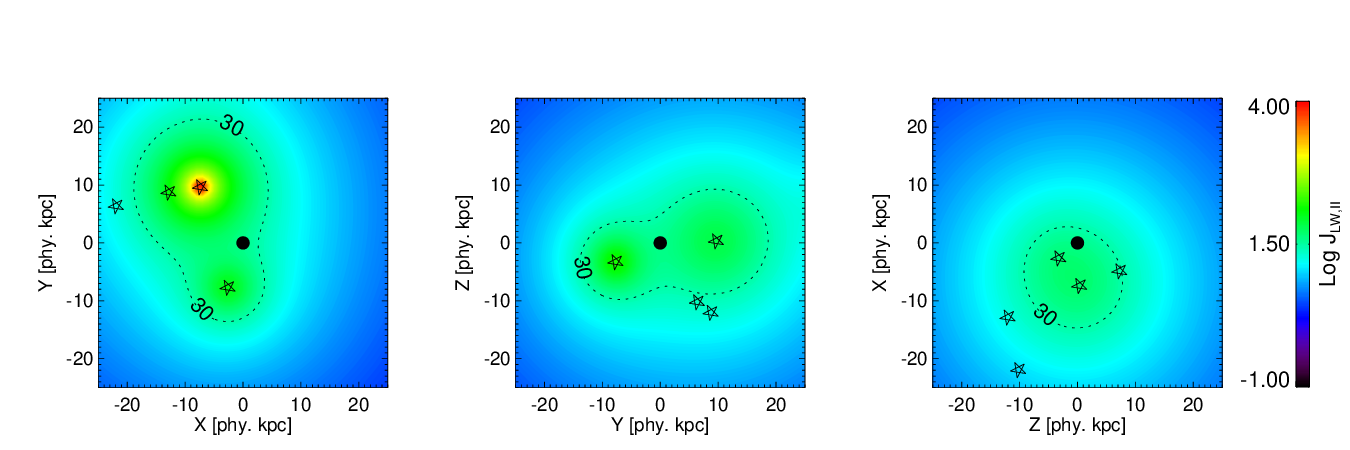}
\includegraphics[trim = 0mm 0mm 0mm 15mm, clip,width=1.75\columnwidth,]{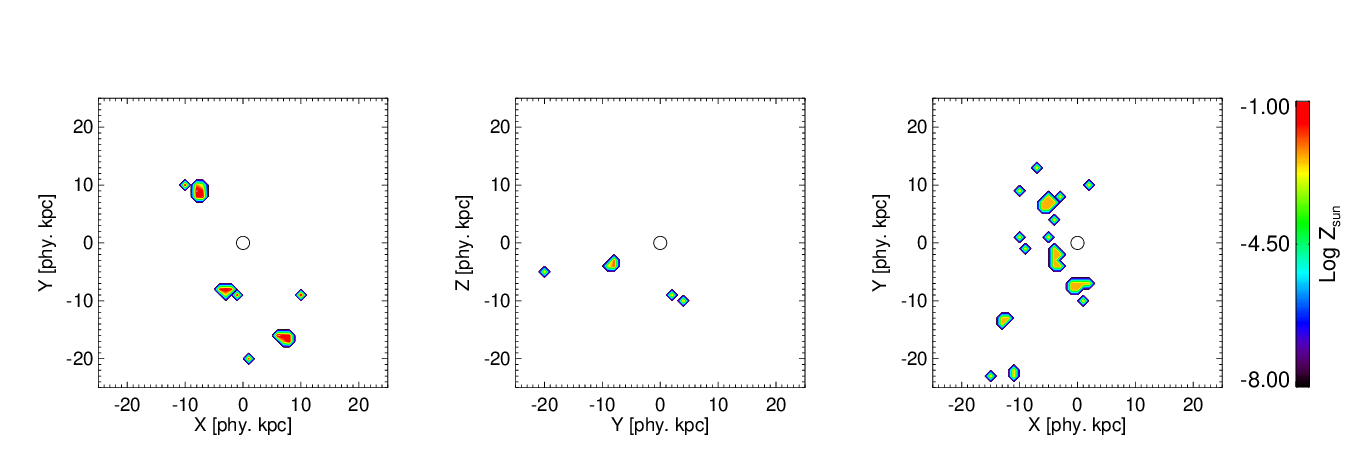}
\caption[LW radiation and metallicity for DC2's environment]{Same as in Fig. \ref{fig.DC0 slices} but for DC2.}
\label{fig.DC2 slices}
\end{figure*}

\begin{figure*}
\centering
\includegraphics[trim = 0mm 0mm 0mm 15mm, clip,width=1.75\columnwidth,]{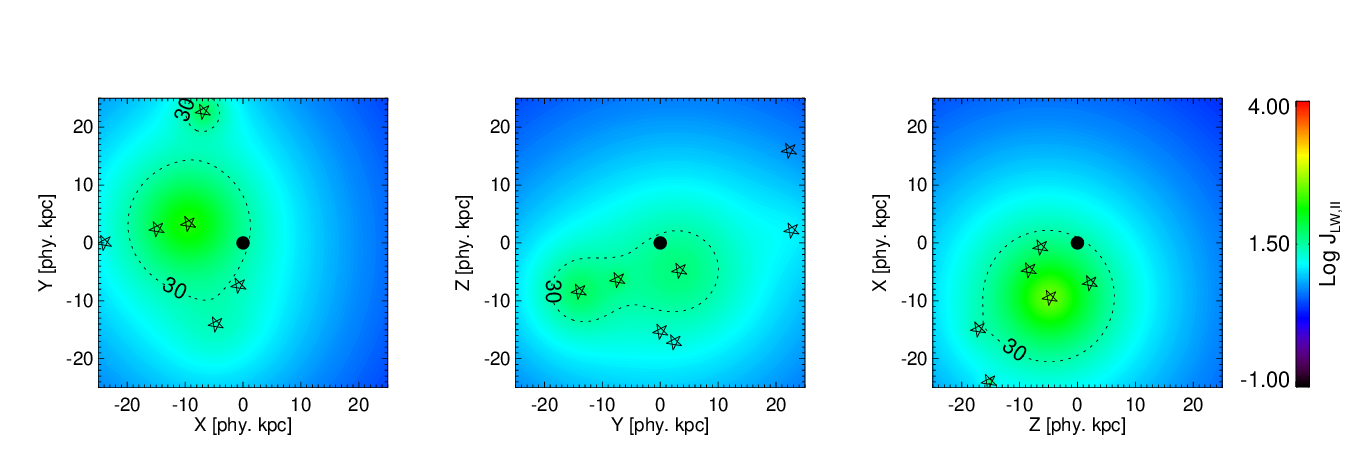}
\includegraphics[trim = 0mm 0mm 0mm 15mm, clip,width=1.75\columnwidth,]{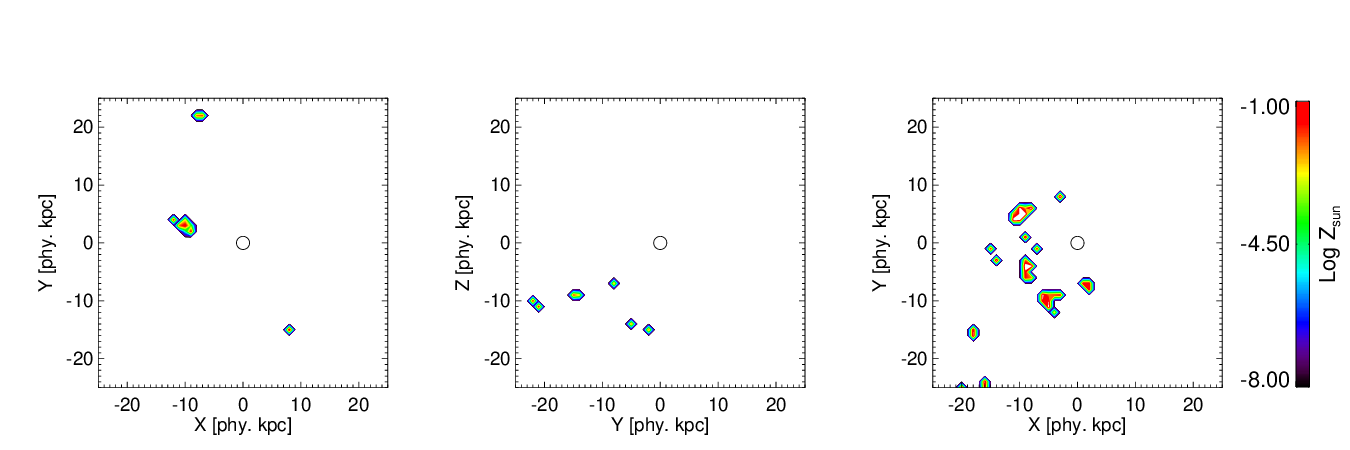}
\caption[LW radiation and metallicity for DC3's environment]{Same as in Fig. \ref{fig.DC0 slices} but for DC3.}
\label{fig.DC3 slices}
\end{figure*}

Given that a critical level of LW radiation needs to be generated by actively star forming galaxies in the vicinity of the candidate halo, DCBHs generally form as satellites \footnote{In this study, we call satellite haloes any (sub)halo within the same \textit{friends-of-friends} group} of a larger galaxy which was giving out a major part of the LW radiation \citep[A12,][A13 hereafter]{Agarwal:2013p2461}. 
In Fig.~\ref{fig.mergertrees}, we plot the merger history of the DCBH haloes, where the main progenitor branch of the halo with which the DCBH host halo merges, is shown to the left, and the halo history of the DCBH halo itself is shown to the right. The DC candidate haloes generally form as satellites of a larger galaxy (except for DC3), with which they eventually merge. This is expected as a larger galaxies' LW flux would be imperative in quenching Pop III SF early on in the DC candidate's history. The implications of the merger trees are discussed in more detail in the following subsections. In the top panels of Fig.~\ref{fig.DC0 slices}-\ref{fig.DC3 slices}, we show the local variation of $J_{\rm LW, II}$ from the neighbouring galaxies and the corresponding metallicity of the gas and stars in the same region is shown in the bottom most panel. The DCBH host haloes always lie inside the contour that marks $J_{\rm crit, II}=30$, where the contour is either the resultant of one \textit{dominant} galaxy or cumulative of various galaxies. The corresponding metallicity slices accurately depict that the DCBH host haloes in question are metal free, and the same galaxies producing the LW flux that allow for DCBH formation in a candidate halo do not necessarily pollute them with metals. This is the first time that the issue of pristine atomic cooling haloes simultaneously undergoing metal pollution and exposure to LW radiation has been addressed. On the basis of the above plots, three distinct scenarios in which a halo can host a DCBH site emerge. We describe them in the following sections.

\subsubsection{Case 1}
\begin{itemize}
\item \textit{Formation in a site close to one dominant galaxy and subsequent merging with it} (DC0, DC1, DC4, DC5). 
\end{itemize}

In this case (see left panel Fig. \ref{fig.mergertrees}), the DC halo virialises near a proto-galaxy (M$_* \approx 10^7 \msun$) that formed early in the simulation. The galaxy has already had a prior episode of Pop III star formation that led to an onset of Pop II star formation. The LW flux from the galaxy is high enough to quench star formation in its vicinity. The quenching of star formation in the vicinity of the large galaxy proves extremely beneficial to the existence of a DC candidate halo later on. The quenching of star formation makes certain that no Pop III stars form in pristine minihaloes and explode as SNe later on to pollute the DC candidate halo. The DC candidate halo eventually merges with the larger galaxy a few time steps later. An interesting outcome is that the DCBH seeds do not necessarily form in the (larger) galaxies that eventually end up hosting them. The DC candidate halo does not grow much before encountering the merger event, thus it is unlikely that the DCBH would grow considerably before the merger event.

\subsubsection{Case 2}
\begin{itemize}
\item \textit{Formation in a clustered environment with subsequent merging with one of the nearby sources} (DC2).
\end{itemize}

The critical level of LW radiation is a combined effect of various low mass proto-galaxies (M$_* \approx 10^{5-7} \msun$) close to the DC halo. Some of the galaxies could also be the satellites of a larger galaxy. In this case both the DC halo and the larger galaxy's host haloes virialise quite early on (with $T_{\rm vir}\sim 2000\ \rm K$), almost at the same epoch (see middle panel Fig.~\ref{fig.mergertrees}), but the larger galaxy's halo is considerably larger which allows it to first host Pop III stars, followed by Pop II star formation. However, it appears that the larger galaxy in this case is not able to produce the critical LW flux by itself and is aided by other galaxies (top panel Fig.~\ref{fig.DC2 slices}). We will further address the issue of what galactic neighbourhood is able to produce a critical LW flux in the section~\ref{stellar neighbourhood}. Similar to case~1, the DC candidate halo merges with a larger galaxy, however, it does so later on as compared to case~1. The DC candidate halo itself grows to $\approx 10^8 \msun$, which could provide enough fuel for subsequent gas accretion by the DCBH before it merges with the larger galaxy.

\subsubsection{Case 3}
\begin{itemize}
\item \textit{Formation in clustered environment without subsequent merging} (see DC3). 
\end{itemize}
DCBH host haloes are generally the satellites of a galaxy. However, DC3 forms in a stand--alone main progenitor branch, without being a satellite of any associated LW flux producing galaxy (see right panel Fig.~\ref{fig.mergertrees}). The top panels of Fig.~\ref{fig.DC3 slices} reveal that there are other galaxies in the neighbourhood (M$_* \approx 10^{5-7} \msun$) that are giving rise to the critical LW flux. These galaxies operate in a similar fashion as described in case~2 by quenching star formation early on with a subcritical LW flux \citep{Machacek:2001p150,OShea:2008p41}, and ultimately producing the critical LW flux at a time when the DC candidate halo crosses the atomic cooling limit. The DC halo might still merge with a larger halo at later times similar as in case~2, but did not do so by $z=6$, when our simulation ends.

\subsubsection*{}
Investigating the level of LW radiation and metal pollution in the DC candidate haloes' neighbourhood shows that the same galaxies that produce the critical level of LW specific intensity, and therefore are in close vicinity of the pristine host halo, do not necessarily pollute the aforementioned haloes with metals at the instant when they are ideal sites for DC. The haloes constituting the sample DC0-DC5 have been identified at an epoch when they just cross the atomic cooling threshold and the process of DC will ensue at this point. Given that the timescale for DCBH formation is similar to that of Pop~III star formation, i.e. few Myr, \citep{Latif:2013p2787,Regan:2009p776,Wise:2008p789}, it is further unlikely that the neighbouring galaxies will pollute these candidate haloes with metals and inhibit DCBH formation. 


\subsection{Galaxies producing $J_{\rm crit}$}
\label{stellar neighbourhood}
\begin{figure}
\centering
\includegraphics[width=\columnwidth,]{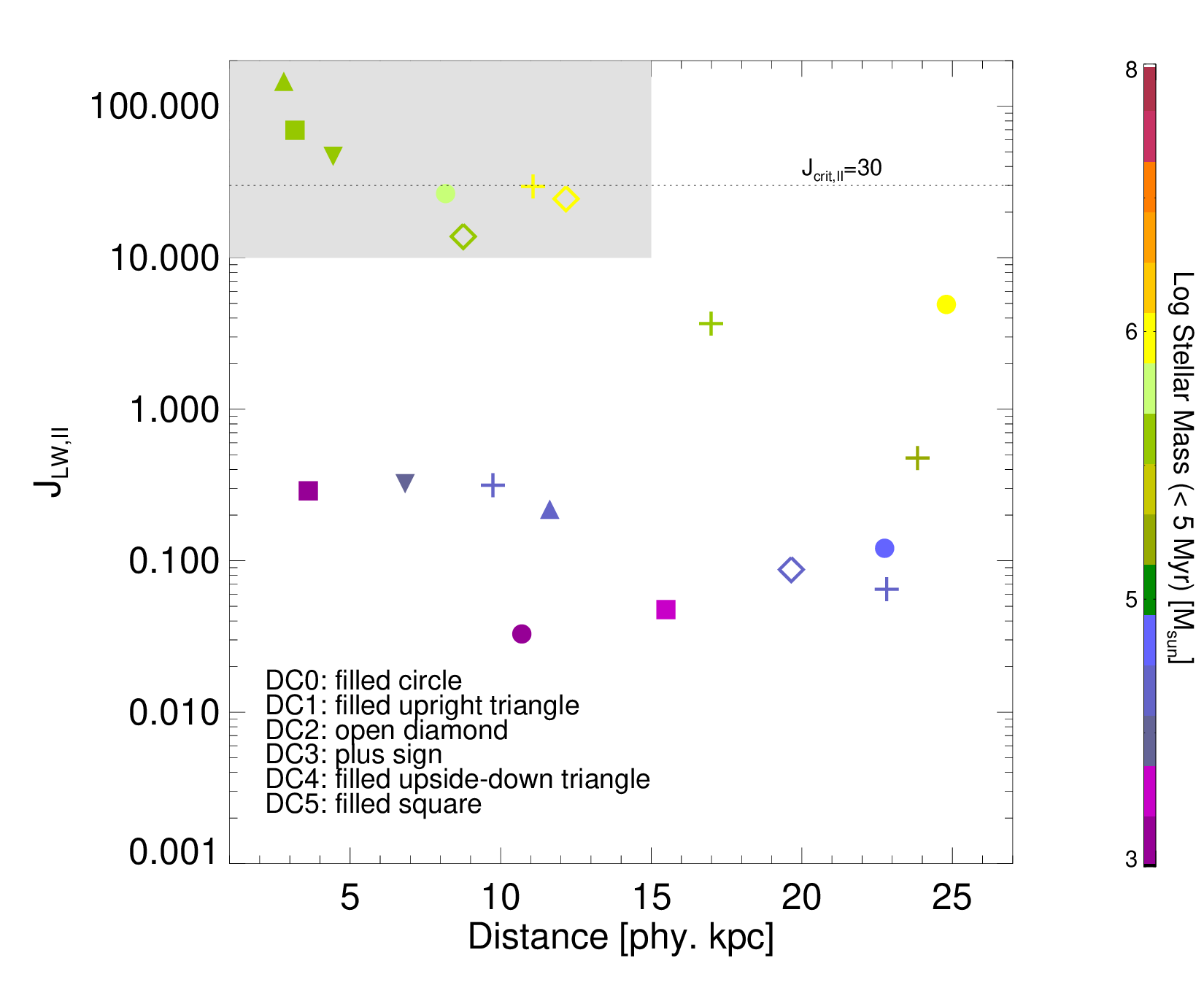}
\includegraphics[width=\columnwidth,]{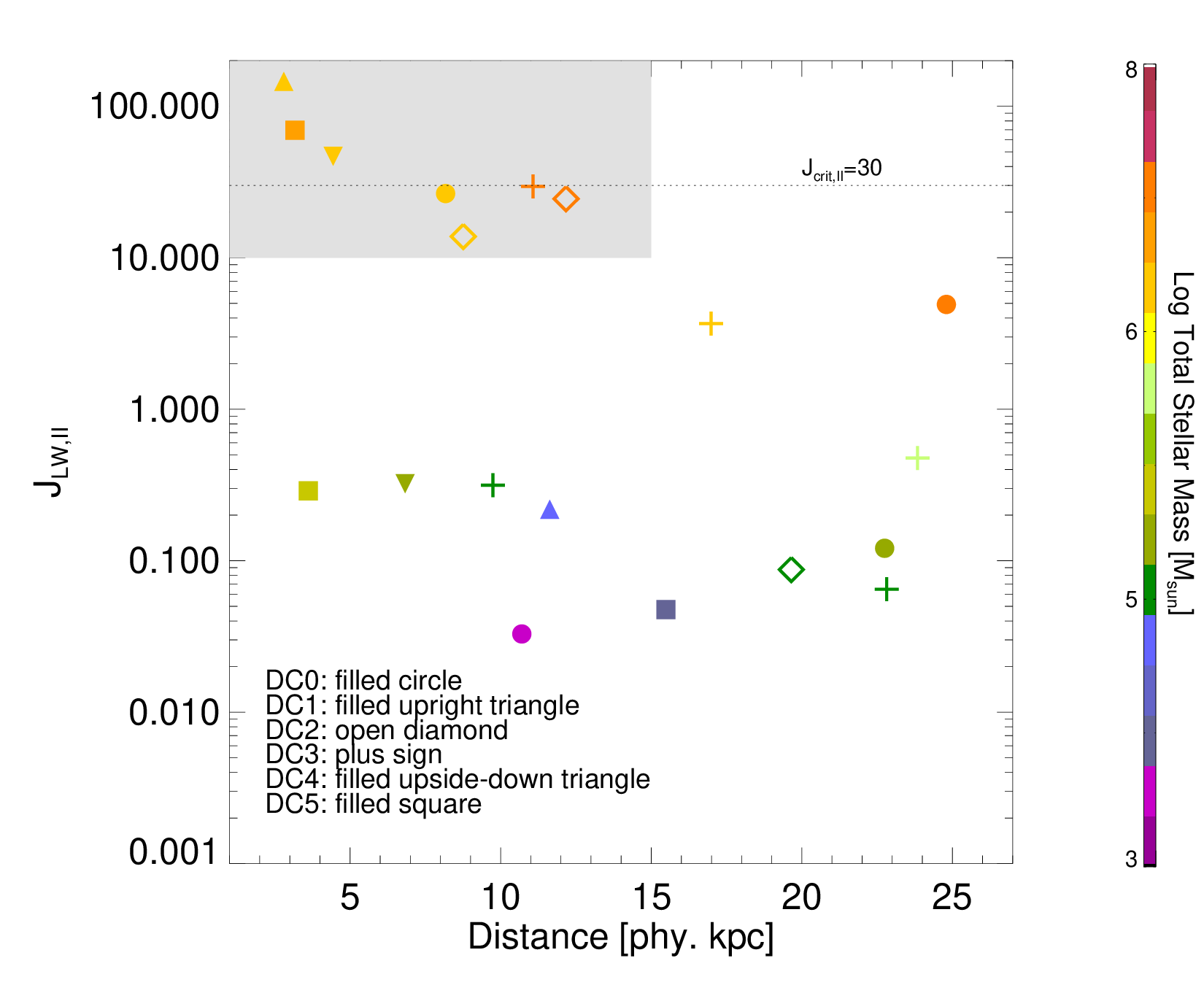}
\caption[Galaxies in the neighbourhood of the DC candidates]{The LW specific intensity produced by each of the neighbouring galaxies at a given distance as seen in the LW contour plots of Fig. \ref{fig.DC0 slices}-\ref{fig.DC3 slices}. Each set of symbols corresponds to a DC candidate, where the symbols represent a galaxy found in the field. We colour code the galaxies by their stellar mass that is $<5$ Myr old (top panel) and the total stellar mass (bottom panel).}
\label{fig.scatter}
\end{figure}

In Fig.~\ref{fig.scatter}, we plot the LW specific intensity versus the physical distance of each of the galaxies producing it, as seen in Fig.~\ref{fig.DC0 slices}-\ref{fig.DC3 slices}. Each symbol style represents a DC case, where the symbols correspond to the galaxies found in the local neighbourhood of the DC site.
In order to understand the nature of the stellar population in each galaxy that gives rise to $J_{\rm LW, II}^{\rm local}$, we colour code the symbols by the amount of stellar mass formed within $5\Myr$ in the top panel and the total stellar mass of the same galaxies in the bottom panel.

The proto-galaxies that produce $J_{\rm LW, II}\geq 10$ have formed at least $\sim 5\times 10^5\msun$ in stars within the past $5\Myr$, i.e.~a SFR of at least $\gtsim 0.1\msunyr$, and are also predominantly composed of Pop~II stars. For all the DC sites, at least one proto-galaxy with a stellar mass larger than $10^6\msun$ is found at a distance of $d \leq 15\kpc$. The grey region bounds these two limits of $J_{\rm LW, II} = 10$ and $d=15\kpc$. The proto-galaxies that lie in the grey region are the ones that contribute either solely (most noticeably: DC1, DC4, DC5, i.e. filled up right triangle, filled upside down triangle and filled square) or cumulatively (most noticeably: DC2, i.e. open diamonds) to $J_{\rm crit, II}$, represented by the dotted line in each of the panels. The case of DC2 and DC3 here denote how multiple galaxies, spread across the shaded grey and unshaded region together contribute towards the critical flux. It is a combination of the distance and the amount of star formation in the neighbouring galaxies for these two haloes, that leads to its exposure to the critical flux. This is contrary to the other cases where a single close by galaxy is able to produce the critical LW flux, unaided by other sources. The galaxies producing the critical LW flux in such cases, all lie in the grey region.

\section{Summary and Discussion}
\label{sec.summary}

In this study, we have employed one of the FiBY project's simulations to pin-point the location and environment of metal-free, non star forming, atomic cooling haloes within a cosmological hydrodynamical simulation. We report a LW flux that is considerably higher than the global mean in 6 candidates, which makes them potential candidates for DCBH formation sites. Our attempt was to quantify if such sites could exist in a cosmological simulation that forms Pop III and Pop II stars self-consistently, accounts for a self--consistent build up of local and global LW radiation flux, and includes metal dispersion via SNe and stellar winds. In order to identify such DC sites, we first identified a sample of pristine, atomic cooling, non star forming haloes and then selected the ones that are exposed to the highest levels of LW radiation, as outlined in Sec. \ref{sec.dc selection}.

The sample of 6 haloes identified in this study hints towards the haloes being possible sites of DCBH, however, the formation of a DCBH would depend on the state of the subsequent gas collapse, which could be probed by extracting these haloes and simulating them in a zoom hydrodynamical simulation that has a high enough resolution. The fact that we have a handful of potential DC sites in our 4 cMpc side-box suggest that DCBHs do not need high-$\sigma$ regions to form, and in fact, can even form in stand alone haloes that happen to be in the vicinity of a few modestly star forming galaxies that cumulatively produce the critical level of LW radiation (see DC3, Fig.\ref{fig.mergertrees}). We report that satellite haloes of larger galaxies are the most likely sites for DCBH formation, i.e. massive seed BHs form outside the galaxies they eventually end up in (A12, A13).

The critical value of the LW specific intensity that favours the formation of DCBHs has been derived in the literature by studying haloes in isolation, where an ideal atomic-cooling halo is selected from a cosmological setup and is irradiated with an increasing level of LW flux till a point is reached where H$_2$ cooling becomes insufficient \citep{Bromm:2003p22,Shang:2010p33,WolcottGreen:2011p121}. This is an assumption on the physical conditions, as the halo under question would be subject to time varying LW feedback from neighbouring galaxies right since its birth. Therefore the previous calculations of $J_{\rm crit}$ might have been overestimated. The exposure of the halo to a time--varying LW flux, ever since its birth, was self-consistently accounted for in the FiBY simulation analysed in this study. \footnote{However note that the authors used reaction rates based on a black body spectrum with a temperature of $10^4$ and $10^5\ \rm K$ representative of a Pop II and Pop III stellar population respectively.}
Therefore whether or not the rest of the haloes (besides the 6 DC candidates) in the sample that are exposed to lower values of $J_{\rm LW}$ could harbour a DCBH is unclear, as the only criterion the other haloes do not meet is the exposure to high levels of LW radiation.

The subsequent accretion process and the final mass that these DCBHs attain would be highly dependent on the mergers that the DC haloes go through. As mentioned earlier, DC0, DC1, DC4, DC5 form as satellites of a larger galaxies and eventually undergo mergers. Upon formation, the DCBH could engulf a major fraction of the gas in its host galaxy \citep[see for e.g.][]{Schleicher:2013p3661}, thereby running out of gas for subsequent accretion. Mergers with larger gas rich galaxies could turn on the accretion process again, aiding these DCBHs to attain supermassive scales (A13). 
Surprisingly, DC3 does not form as a satellite of a larger galaxy and the host halo attains a mass of $\approx 10^{7.5}\msun$ by the end of the simulation ($z=6$). Whether this particular candidate evolves into a Milky Way type galaxy, or ends up in a scenario of quenched DCBH accretion due to insufficient fuelling is unclear and the subject of an undergoing study.

Note that DC4 and DC5 end up in the same galaxy at $z=6$, hinting towards the possible event of a DCBH merger in the early Universe. 
This sort of event, would be an ideal candidate in explaining the growth of massive seed BHs to supermassive scales, where upon undergoing a merger, the seeds could double their mass and continue to grow by rapid gas accretion. However, mergers are subject to gravitational recoil and dynamical friction effects, which could hinder the growth of these DCBHs. In order to find such an optimised event, one would need to run our simulation with a much larger box size, a feat unattainable with the current computational limitations. That said, the occurrence of these sites in our relatively-small simulation volume hints towards the possibility that most present day galaxies might be harbouring a DCBH at their centres. 

In this study we haven't touched upon the effects of reionisation on the DCBH formation sites. The candidate haloes cross the atomic cooling limit at redshifts between 8 and 10, when the CMB data suggests that reionisation was already under way \citep{Komatsu:2011p409,Ade:2013p2565}. Moreover, the local sources that produce the high level of LW radiation at the potential formation sites likely also produce a high level of ionising radiation. Thus, it is possible that reionisation affects the candidate DCBH sites. The effect of reionisation on DCBH formation is unclear. Ionising photons could heat the gas in haloes that are unable to shield from the radiation, preventing the haloes from growing and inhibiting collapse to a DCBH. Also, the additional free electrons allow for faster H$_2$ formation through the H$^-$ channel at low densities, causing the gas inside the halo to cool more efficiently. Since the effect of reionisation on DCBH formation depends critically on the ability of the formation sites to self-shield against the ionising radiation, addressing this issue requires simulations with a higher resolution than we employ here. We have therefore decided to address these issues in a follow-up study \citep{Johnson:2014p3558}.

The next step is to extract these haloes and simulate them for their entire formation histories, with the associated LW radiation (and other properties) as input, in high-resolution zoom simulations. This will shed new light on the role that LW radiation and an ionising flux, amongst other properties, plays on the process of DCBH formation.


\section*{Acknowledgements}
The simulation was run on the facilities of the Max-Planck Rechenzentrum Garching.
BA would like to thank Andrew Davis for his extremely useful inputs during the early stages of the study. BA would also like to thank Jonny Elliott and Alessia Longobardi for their comments during the preparation of the manuscript.
CDV acknowledges support by Marie Curie Reintegration Grant PERG06-GA-2009-256573.
Work at LANL was done under the auspices of the National Nuclear Security Administration of the US Department of Energy at Los Alamos National Laboratory under Contract No. DE-AC52-06NA25396.
 \bibliographystyle{mn2e}
 \bibliography{babib}
\label{lastpage}

\end{document}